\documentclass[11pt,a4paper]{article}
\usepackage{amsmath}
\usepackage{amssymb}
\usepackage{amsfonts}

\begin{document}

\pagestyle{myheadings}

\begin{center}
\noindent{\bf Natural Numbers and Quantum States in Fock Space}\\
\vspace{5mm}
Francesco A. Raffa$\,^1$ and Mario Rasetti$\,^2$\\
\vspace{5mm}
$^1$ Dipartimento di Meccanica, Politecnico di Torino, 10129 Torino, Italy\\
francesco.raffa@polito.it\\
$^2$ Dipartimento di Fisica, Politecnico di Torino, 10129 Torino, Italy\\
mario.rasetti@polito.it
\end{center}

\vspace{15mm}

\noindent{\bf Abstract}\\
We investigate the expression of natural numbers in any base from a quantum point of view. In particular, resorting to the one-to-one correspondence between natural numbers and Fock states, we construct a set of multiboson operators and a set of translation operators, whose action on the Fock states leads to the coefficients identifying a natural number in any base.


\section{Introduction}
The importance of natural numbers in both the theoretical foundations and the experimental reality of quantum mechanics is widely recognized. This seems to be especially true in the field of quantum computation and quantum information theory, the latter including the remarkable instrument of quantum cryptography \cite{NIECHU}. There are therefore both speculative and practical reasons to examine the connection between natural numbers and quantum number states, as well as the operations one can define on them.

This topic has been addressed in Ref. \cite{BENIOFF1}, where the representation of natural numbers through tensor product states and the basic arithmetic operations of addition and multiplication are investigated utilizing the axiomatic description of numbers and resorting to an abstract and a physical Hilbert space connected by unitary maps. A further approach is reported in Refs. \cite{BENIOFF2}, \cite{BENIOFF3} where natural, integer and rational numbers are represented as states of finite strings of base $k$ qukits, $k \geq 2$, located on a two-dimensional integer lattice. In Ref. \cite{BENIOFF3}, in particular, the theoretical features of the translation and base change operations are discussed.

The quantum representations of numbers and operations are analyzed in Benioff's papers from a fundamental, formal point of view; however, scarce attention is devoted to the realization of the relevant operators. An example of explicit realization is reported in Ref. \cite{RARAijtp} where, with reference to spin $\frac{1}{2}$ systems corresponding to the binary base, fermionized translation operators are constructed, which satisfy the anticommutation relations through appropriate phase factor operators.

A different problem is investigated in this paper. Specifically, we consider the possibility of defining quantum operators which calculate the coefficients of the expansion of a natural number in any base and report two different solutions of this problem. Both are based on the one-to-one correspondence between natural numbers and number states in the Fock space, though their realizations are distinct: indeed, one of them entails the definition of specific multiboson operators, while the other one utilizes translation operators which, unlike those reported in Ref. \cite{RARAijtp}, are not endowed with a definite algebraic structure.

The organization of the paper is as follows: preliminary definitions concerning the finite and the infinite register and the corresponding structure of the Fock space are given in section 2. In section 3, in the framework of the multiboson algebras, a new representation of the multiboson number operator is obtained resorting to the $z$-transform method; such representation enters the definition of both the multiboson operators and the translation operators relevant to this paper. The construction and the properties of the translation operators are illustrated in section 4, while in section 5 both sets of operators are applied to the number state to extract the coefficients of the corresponding natural number. 

\section{The Registers and the Fock Space}
For a finite $q$-slot register the natural number $n$ in any base $b \geq 2$ is written as
\begin{equation}
n \,=\, \sum_{\ell =0}^{q-1} \gamma_{\ell} \, b^{\ell} \; , \label{NUMBASEB}
\end{equation}
where $q \geq 1$, $0 \leq n \leq N$, $N$ $=$ $b^q - 1$, $\gamma_\ell$ $\in$ $\{0,1, \ldots, x\}$, with $x$ $=$ $b-1$. The corresponding number state $|n\rangle$ lives in the subspace ${\mathfrak{F}}_q \doteq {\rm span}\{|n\rangle \, | \, 0 \leq n \leq N \}$ of the Fock space ${\mathfrak{F}} \doteq {\rm span} \{ |n\rangle \, | \, n \in \mathbb{N} \}$. Notice that $|n\rangle$ can be written also in the tensor product form
\begin{equation}
|n\rangle \,=\, |\gamma_0\rangle \otimes |\gamma_1\rangle \otimes \cdots \otimes |\gamma_\ell\rangle \otimes \cdots \otimes |\gamma_{q-1}\rangle \; , \label{NUMBERTENSOR}
\end{equation}
from which, since $|\gamma_\ell\rangle$ $\in$ ${\mathfrak{H}}_\ell \sim {\mathbb{C}}^b$, it follows that ${\mathfrak{F}}_q$ $=$ $\displaystyle \bigotimes_{\ell = 0}^{q-1} {\mathfrak{H}}_\ell$ $\sim$ $({\mathbb{C}}^b)^{\otimes q}$. For a register with an infinite number of slots one has
\begin{equation}
n \,=\, \sum_{\ell =0}^{\infty} \gamma_{\ell} \, b^{\ell} \; , \label{NUMBASEBinf}
\end{equation}
while the number state $|n\rangle$, written either in the standard or in the tensor product form, is an element of the whole Fock space $\mathfrak{F}$. 

\section{The Multiboson Number Operator}
The $k$-boson algebras $\displaystyle \{ {\mathbb{I}}, A_k, A_k^{\dagger}, \hat{N}_k \,=\, A_k^{\dagger} A_k \}$, $k$ integer $\geq$ 1, introduced in Ref. \cite{BRANDT}, are a generalization of the single-boson algebra $\displaystyle \{ {\mathbb{I}}, a, a^{\dagger}, {\hat{n}} \,=\, a^{\dagger} a \}$. $\displaystyle {\hat{N}}_k$ is the multiboson number operator, while $A_k$ and $A_k^{\dagger}$ are designed to annihilate and create $k$ bosons at a time, $\displaystyle \left[A_k,\hat{n}\right]$ $=$ $k \, A_k$, $\displaystyle [A_k^{\dagger},\hat{n}]$ $=$ $- k \, A_k^{\dagger}$. The multiboson operators obey the canonical commutation relations $\displaystyle \left[ \mathbb{I}, \bullet \right]$ $=$ 0, $\displaystyle [A_k, A_k^{\dagger}]$ $=$ ${\mathbb{I}}$, $\displaystyle [ {\hat{N}}_k, A_k ]$ $=$ $- A_k$. One defines also the operator 
\begin{equation}
{\hat{D}}_k \,=\, {\hat{n}} - k \, {\hat{N}}_k \; , \label{DK}
\end{equation}
which has no counterpart in the single-boson algebra.

It is quite natural that $k$-boson operators, by their very definition, can be expressed in terms of single-boson operators. In Ref. \cite{BRANDT} the following power series representation of $A_k$ in $a$, $a^{\dagger}$ is obtained
\begin{equation}
A_k \,=\, \sum_{j=0}^{\infty} \alpha_{j}^{(k)} {a^{\dagger}}^j a^{j+k} \quad,\quad \alpha_{j}^{(k)} \,=\, \sum_{l=0}^{j} \frac{(-)^{j - l}}{(j-l)!} \left (\frac{1+\left\lfloor \frac{l}{k} \right\rfloor}{l!(l+k)!} \right )^{\frac{1}{2}} \; , \label{AK-BRANDT}
\end{equation}
where the symbol $\displaystyle \lfloor y \rfloor$ denotes the integral part of $y$, i.e., the maximum integer $\leq y$. In view of Eq. \eqref{AK-BRANDT}, $\displaystyle \hat{N}_k$ proves to be the sum of an infinite number of polynomials of degree $j+m+k$ in $\hat{n}$
\begin{equation}
\hat{N}_k \,=\, \sum^{\infty}_{j,m=0} \alpha_{m}^{(k)} \alpha_{j}^{(k)} \sum^{j+k}_{r=0} \sum^{m}_{s=0} S_{j+k} ^{(r)} \, S_{m} ^{(s)} \, {\hat{n}}^r \left( {\hat{n}-k} \right)^s \; , \label{NKPOL}
\end{equation}
where the coefficients $S_{j+k} ^{(r)}$, $S_{m} ^{(s)}$ are Stirling numbers of the first kind \cite{ABRAM}. An alternative representation of $A_k$ is reported in Ref. \cite{RASETTI}
\begin{equation}
A_k \,=\, a^k \, F_k(\hat{n}) \quad,\quad F_k(\hat{n}) \,=\, \left( \left\lfloor \frac{\hat{n}}{k} \right\rfloor \frac{(\hat{n}-k)!} {\hat{n}!} \right)^{\frac{1}{2}} \; .\label{AK-RASET}
\end{equation}
Notice that Eq. (\ref{AK-RASET}) gives for the multiboson number operator the interpretation 
\begin{equation}
\hat{N}_k \,=\, \left\lfloor \frac{\hat{n}}{k} \right\rfloor \; , \label{NKpartint}
\end{equation}
which identifies $\displaystyle {\hat{D}}_k$ as the remainder of $\hat{n}$ mod $k$: indeed, with $n$ in modular arithmetic, $n$ $=$ $\displaystyle k \left\lfloor \frac{n}{k} \right\rfloor$ $+$ $t$, $t$ $\in$ $\{0,1, \ldots, k-1\}$, one has $\displaystyle \hat{D}_k |n\rangle$ $=$ $t |n\rangle$.

It is also worth noticing that since the known representations of the multiboson number operator require the use of the integral part of rationals, either through the coefficients $\displaystyle \alpha_{j}^{(k)}$ in Eq. \eqref{AK-BRANDT} or in the very definition (\ref{NKpartint}) of the operator, their use is in some way a source of logical inconsistency because the integral part appears in the operator, besides being its eigenvalue.

A representation of $\hat{N}_k$, which does not contain the integral part, can be found considering the eigenvalue equation for $\hat{N}_k$ in the Fock space ${\mathfrak{F}}$. In terms of $n$ the eigenvalue $\displaystyle \left\lfloor \frac{n}{k} \right\rfloor$ is the sum of infinite blocks, each completely filled with $k$ discrete pulses centered at the points $n$ $=$ $\ell k + m$, $m$ $=$ $0, 1, \ldots, k-1$, with $\ell$ the amplitude of the pulses in the $(\ell + 1)$-th block,
\begin{equation}
\left\lfloor \frac{n}{k} \right\rfloor \,=\, \sum^{\infty}_{\ell=0} \ell \sum^{k-1}_{m=0} \delta[n - (\ell k + m)] \label{Z-SIGNAL} \; ,
\end{equation}
$\delta[n-y]$ denoting the discrete pulse of unit amplitude centered at $y$. The $z$-transform of the function (\ref{Z-SIGNAL}), with $z$ $\in$ $\mathbb{C}$, is the Laurent series $\displaystyle \sum^{\infty}_{n=0} \left\lfloor \frac{n}{k} \right\rfloor \, z^{-n}$ \cite{JURY}, whose sum, for $|z| > 1$, is $\displaystyle {\cal N}(z)$ $=$ $\displaystyle z \, (z-1)^{-1} (z^k-1)^{-1}$. The function (\ref{Z-SIGNAL}) is restored evaluating the inverse $z$-transform of $\displaystyle {\cal N}(z)$ through the complex integral formula \cite{JURY} $\displaystyle \left\lfloor \frac{n}{k} \right\rfloor$ $=$ $\displaystyle \frac{1}{2 \pi i} \oint_{\Gamma} {\cal N}(z) z^{n-1} dz$, where $\Gamma$ is a counterclockwise path of integration enclosing the origin and lying in the region of the $z$-plane in which ${\cal N}(z)$ is holomorphic, $|z| > 1$. Replacing $n$ with $\hat{n}$ leads to
\begin{equation}
\left\lfloor \frac{\hat{n}}{k} \right\rfloor \,=\, \frac{1}{2 \pi i} \oint_{\Gamma} \frac{z^{\hat{n}}}{\left(z-1\right) \left(z^k-1\right)} dz \; . \label{NK-ZETA-INF}
\end{equation}

The analytic evaluation of the contour integral in Eq. (\ref{NK-ZETA-INF}) is carried out introducing the polynomial $\displaystyle P^{(k-1)}(z)$ of degree $k-1$ in $z$,
\begin{equation}
P^{(k-1)}(z) \,=\, \sum^{k-1}_{\ell=0} z^\ell \,=\, \prod^{k-1}_{\ell=1} \left(z - {\zeta_{\ell}} \right) \; , \label{PKZ}
\end{equation}
such that $\displaystyle z^k - 1$ $=$ $\displaystyle (z-1)P^{(k-1)}(z)$. In Eq. \eqref{PKZ} $\displaystyle {\zeta_{\ell}}$ $=$ $\exp (i 2 \pi \ell/k)$ while $\displaystyle \prod^{0}_{\ell=1}$ $\equiv$ 1 as $P^{(0)}(z)$ $=$ 1. The singularities inside the path of integration are thus seen to be the 2nd-order pole at $z = 1$ and the $k-1$ simple poles at $\displaystyle z =  {\zeta_{\ell}}$, $\ell$ $=$ $1,\ldots,k-1$ (the latter are the $k$-th roots of unity with the exclusion of $z = 1$). Resorting to the Cauchy residue theorem one obtains 
\begin{equation}
\left\lfloor \frac{\hat{n}}{k} \right\rfloor \,=\, \frac{2\hat{n} - k + 1}{2k} + \sum^{k-1}_{j=1} C_j ^{(k)} \, \zeta_j^{\hat{n}} \; , \label{NK-FINAL}
\end{equation}
where $\displaystyle \sum^{0}_{j=1}$ $\equiv$ 0 and the coefficients $\displaystyle C_j ^{(k)}$, for $k \geq 2$, are complex numbers that can be expressed in the form of a rational function of degree $k$ in $\zeta_{j}$
\begin{equation}
C_j ^{(k)} \,=\, \left({\zeta_{j}}-1 \right)^{-2} \, \prod^{k-1}_{{\ell=1}\atop{(\neq j)}} \left( {\zeta_j} - {\zeta_{\ell}} \right)^{-1} \; . \label{QK}
\end{equation}
Notice that the first term in the rhs of Eq. (\ref{NK-FINAL}), produced by the residue at the double pole, is linear in $\hat{n}$ and applies for $k \geq 1$, while the second one, which results from the residues at the simple poles, is nonlinear in $\hat{n}$ and applies for $k \geq 2$.

\section{The Translation Transformation}
In this section the construction of the translation operators is reported progressing from the single slot to the whole register.

\subsection{The slot operators}
Depending on the values of $\gamma_{\ell}$, the action of the lowering and raising operator, $S^{(\ell)}$ and ${S^\dagger} ^{(\ell)}$, on the $(\ell + 1)$-th slot of the number state \eqref{NUMBERTENSOR} is required to be
\begin{eqnarray}
S^{(\ell)} \, |\gamma_{\ell}\rangle &=& |\gamma_{\ell}-1\rangle \;,\; 1 \leq \gamma_{\ell} \leq x \quad,\quad S^{(\ell)} \, |0\rangle \,=\, 0 \label{S-SLOT}\\ 
{S^\dagger} ^{(\ell)} \, |\gamma_{\ell}\rangle &=& |\gamma_{\ell}+1\rangle \;,\; 0 \leq \gamma_{\ell} \leq x-1 \quad,\quad {S^\dagger} ^{(\ell)} \, |x\rangle \,=\, 0 \; . \label{SDAG-SLOT}
\end{eqnarray}

One derives the expression of $\displaystyle S^{(\ell)}$ extending the formulation reported in Ref. \cite{RARAijtp} for spin $\frac{1}{2}$ systems to general spin $j$ systems. Specifically, $S^{(\ell)}$ and ${S^\dagger} ^{(\ell)}$ are expressed in terms of the spin $j$ representations of the generators $J_-$ and $J_+$ $=$ $J_- ^\dagger$ of the $su(2)$ algebra $\{\displaystyle J_+, J_-, J_z\}$, $\displaystyle [J_+,J_-] \,=\, 2 J_z$, $\displaystyle [J_z,J_{\pm}] \,=\, \pm J_\pm$, the standard basis $|j,m\rangle$, $- j \leq m \leq j$, being replaced with the number basis through the Holstein-Primakoff representation of $su(2)$ \cite{MATTIS}. Omitting computational details, one finds that $\displaystyle S ^{(\ell)}$ $=$ $\displaystyle \frac{1}{\sqrt{\hat{\gamma}_{\ell}+1}} \, a_{\ell}$ satisfies both conditions (\ref{S-SLOT}); $a_\ell$ and $\hat{\gamma}_{\ell}$ denote the annihilation and the number operators of the $(\ell + 1)$-th slot, $a_{\ell} |\gamma_{\ell}\rangle$ $=$ $\sqrt{\gamma_{\ell}} |\gamma_{\ell}-1\rangle$, $\hat{\gamma}_{\ell} |\gamma_{\ell}\rangle$ $=$ $\gamma_{\ell} |\gamma_{\ell}\rangle$. However, the raising slot operator $\displaystyle {S^\dagger} ^{(\ell)}$ verifies only the first of conditions (\ref{SDAG-SLOT}) while the second one needs somehow to be forced . To this aim, with $\mathbb{I}_b$ denoting the $b$-dimensional identity, one defines
\begin{equation}
{S^\dagger} ^{(\ell)} \doteq a_{\ell} ^\dagger \frac{1}{\sqrt{\hat{\gamma}_{\ell}+1}} \left( \mathbb{I}_b - \left\lfloor \frac{\hat{\gamma}_{\ell}}{x} \right\rfloor \right) \; ,  \label{NEWDEF1} 
\end{equation}
where $\displaystyle \left\lfloor \frac{\hat{\gamma}_{\ell}}{x} \right\rfloor \, |\gamma_{\ell}\rangle \,=\, \delta_{\gamma_{\ell},x} \, |\gamma_{\ell}\rangle$, $\delta_{\gamma_{\ell},x}$ $=$ $\displaystyle \left\lfloor \frac{\gamma_{\ell}}{x} \right\rfloor$ being the Kronecker delta. The redefinition of the lowering slot operator, which follows from Eq. \eqref{NEWDEF1}, does not modify the action of $S^{(\ell)}$ on $|\gamma_{\ell}\rangle$.
Furthermore, applying the $z$-transform method as in section 2, one finds a representation of the operator $\displaystyle \left\lfloor \frac{\hat{\gamma}_{\ell}}{x} \right\rfloor$. Its eigenvalue in the subspace $\mathfrak{H}_\ell$ can be considered a single pulse of unit amplitude centered at $x$, whose $z$-transform, $\displaystyle \sum^{\infty}_{\gamma_\ell=0} \delta[\gamma_\ell - x] \, z^{-\gamma_\ell}$ $=$ $z^{-x}$, is readily inverted through the complex integral formula. Upon replacing $\displaystyle \gamma_\ell$ with $\displaystyle \hat{\gamma_\ell}$ and setting $z$ $=$ $|z| \exp(i \vartheta)$, one obtains $\displaystyle \left\lfloor \frac{\hat{\gamma_\ell}}{x} \right\rfloor$ $=$ $\displaystyle |z| e^{\hat{\gamma_\ell} - x} \frac{1}{2 \pi} \int^{2\pi}_{0} e^{-i (x - \hat{\gamma_\ell}) \vartheta} d\vartheta$.

\subsection{The translation operators}
Considering a finite register, one defines quite naturally the operators $\displaystyle t_{\ell}$ and $\displaystyle t_{\ell} ^\dagger$,
\begin{equation}
t_{\ell} \,=\, {\mathbb{I}_b}^{\otimes \ell} \otimes S^{(\ell)} \otimes {\mathbb{I}_b}^{\otimes (q-\ell-1)} \; , \label{TL}
\end{equation}
which act non trivially on the $(\ell+1)$-th slot of the number state \eqref{NUMBERTENSOR}
\begin{equation}
t_{\ell} \, |n\rangle \,=\, |n-b^{\ell}\rangle \;,\; \gamma_\ell \neq 0 \quad,\quad t_{\ell} ^\dagger \, |n\rangle \,=\, |n+b^{\ell}\rangle \;,\; \gamma_{\ell} \neq x \; . \label{TT}
\end{equation}
Eq. (\ref{TT}) shows that the subtraction and addition operations cannot be performed whenever the occupation number in the $(\ell+1)$-th slot is $\gamma_{\ell}$ $=$ 0 and $\gamma_{\ell}$ $=$ $x$, respectively, as in these cases $t_{\ell}$ and $t_{\ell} ^\dagger$ annihilate the state $|n\rangle$. To get round these irreversible `out of the register' exits, one defines the subtraction operators
\begin{equation} 
T_m \,=\, t_m + \sum_{k=m}^{q-2} \, \prod_{j=m}^{k} {t_j^{\dagger}} ^x \, t_{k+1} \;,\; 0 \leq m \leq q-2 \;, \quad \quad T_{q-1} \,=\, t_{q-1} \; , \label{TM}
\end{equation}
and $T_m ^\dagger$, $T_{q-1} ^\dagger$ as addition operators. With $0 \leq m \leq q-1$ one has
\begin{equation} 
T_m |n\rangle \,=\, |n-b^m\rangle \;,\; b^m \leq n \leq N \quad,\quad T_m |n\rangle \,=\, 0 \;,\; 0 \leq n < b^m \; , \label{TSUB}
\end{equation}
\begin{equation}  
T_m ^\dagger |n\rangle \,=\, |n+b^m\rangle \;,\; 0 \leq n \leq N - b^m \quad,\quad T_m ^\dagger |n\rangle \,=\, 0 \;,\; N - b^m < n \leq N \label{TADD} \; .
\end{equation}
One verifies that the number state $|n-b^m\rangle$ is obtained in one of the $(q-m)$ different ways embodied in Eq. \eqref{TM}, depending on the actual values of the coefficients $\gamma_{\ell}$. Generally the shift to be realized is $|n\rangle \longrightarrow |n \pm w\rangle$, where $w$ is a ($q$-slot) natural number, $\displaystyle w \,=\, \sum_{r =0}^{q-1} \beta_r \, b^r$ with $\beta_r$ $\in$ $\{0,1, \ldots, x\}$. In view of the first of Eqs. (\ref{TSUB}), (\ref{TADD})
\begin{equation} 
|n-w\rangle \,=\, \prod_{r=0}^{q-1} T_r ^{\beta_r} |n\rangle \,,\, w \leq n \leq N \;,\; |n+w\rangle \,=\, \prod_{r=0}^{q-1} {T_r ^\dagger} ^{\beta_r} |n\rangle \,,\, 0 \leq n \leq N-w \; . \label{TCONT1}
\end{equation}
Eq. (\ref{TCONT1}) involves the use of all the translation operators. Manifestly, the states $|n \pm w\rangle$ can be obtained resorting only to the operators $T_0$, $T_0 ^\dagger$
\begin{equation} 
|n-w\rangle \,=\, T_0 ^w |n\rangle \;,\; w \leq n \leq N \quad,\quad |n+w\rangle \,=\, {T_0 ^\dagger} ^w |n\rangle \;,\; 0 \leq n \leq N-w \; . \label{TCONT2}
\end{equation}

Minor changes are required for the infinite register: Eq. \eqref{TM} reduces to
\begin{equation} 
T_m \,=\, t_m + \sum_{k=m}^{\infty} \, \prod_{j=m}^{k} {t_j^{\dagger}} ^x \, t_{k+1} \;,\; m \,=\, 0, 1, \ldots
\; , \label{TMINF} 
\end{equation}
while the first of Eqs. (\ref{TSUB}), (\ref{TADD}) are replaced by $\displaystyle T_m |n\rangle$ $=$ $\displaystyle |n-b^m\rangle$, $n \geq b^m$, $\displaystyle T_m ^\dagger |n\rangle$ $=$ $\displaystyle |n+b^m\rangle$, $n \geq 0$. Moreover, the second of Eq. \eqref{TSUB} remains valid, while the second of Eq. \eqref{TADD} does not apply. As for the states $|n \pm w\rangle$, both Eqs. (\ref{TCONT1}) and (\ref{TCONT2}) hold with $n \geq w$ and $n \geq 0$ for the subtraction and addition operations, respectively.

\subsection{Unitarity ranges of the translation operators}
For a finite register Eqs. (\ref{TSUB}), (\ref{TADD}) show that, with respect to the action on the number states, $T_m$ and $T_m ^\dagger$ are unitary, $\displaystyle [T_m,T_m ^\dagger] \,=\, 0$, for $b^m$ $\leq$ $n$ $\leq$ $N - b^m$. The unitarity range has a non zero extension for the physically meaningful values of $b$ and $q$ ($b \geq 2$, $q \geq 1$) with the only exclusion of the case $b = 2$, $m = q-1$.

The left and right vacuum of the register, LV $\doteq$ ${\rm span} \{|n\rangle \, | \, 0 \leq n \leq b^m - 1 \}$ and RV $\doteq$ ${\rm span} \{|n\rangle \, | \, b^q - b^m \leq n \leq N \}$, respectively, are nonunitarity regions. Specifically, in view of Eqs. (\ref{TSUB}), (\ref{TADD}), one verifies that for $|n\rangle$ $\in$ LV, $T_m |n\rangle$ $=$ 0, $T_m ^\dagger |n\rangle$ $=$ $|n+b^m\rangle$, so that $\displaystyle [T_m,T_m ^\dagger] \,=\, {\mathbb{I}}$, while, for $|n\rangle$ $\in$ RV, $T_m |n\rangle$ $=$ $|n-b^m\rangle$, $T_m ^\dagger |n\rangle$ $=$ 0, from which $\displaystyle [T_m,T_m ^\dagger] \,=\, - {\mathbb{I}}$. Notice that for $m = 0$ one obtains the largest unitarity region of the register, as LV and RV reduce to the single states $|0\rangle$ and $|N\rangle$; on the other side, for $m \,=\, q-1$, both LV and RV contain $b^{q-1}$ states, so that the whole register is just the sum of the two vacua when $b = 2$.

For the infinite register the unitarity range enlarges to $n \geq b^m$, the left vacuum of the register being the only nonunitarity region. 

\section{Expansion of $n$ in any Base}
The coefficients $\gamma_{\ell}$ corresponding to the expansion of $n$ in the base $b$ can now be identified as follows.

\subsection{The $\gamma_\ell$'s via multiboson operators}
The $\gamma_{\ell}$'s are calculated resorting to an appropriate generalization of the operator \eqref{DK}. With $n$ as in Eqs. \eqref{NUMBASEB} or \eqref{NUMBASEBinf}, one calculates $\displaystyle \hat{D}_{b^\ell} \, |n\rangle$ $=$ $\displaystyle (\gamma_0 + \gamma_1 b + \ldots + \gamma_{\ell-1} b^{\ell-1}) \, |n\rangle$ as the sum $\displaystyle \gamma_0 b^{-\ell} + \gamma_1 b^{-(\ell-1)} + \ldots + \gamma_{\ell-1} b^{-1}$ does not contribute to $\displaystyle \left\lfloor \frac{n}{b^\ell} \right\rfloor$ for both the finite and the infinite register. Letting the operator
\begin{equation}
\hat{D}_{b} ^{(\ell)} \doteq \frac{1}{b^{\ell}} \left( \hat{D}_{b^{\ell+1}} - \hat{D}_{b^{\ell}} \right) \,=\, \left\lfloor \frac{\hat{n}}{b^\ell} \right\rfloor - b \left\lfloor \frac{\hat{n}}{b^{\ell+1}} \right\rfloor \; , \label{Db}
\end{equation}
act on $|n\rangle$, one finds that $\gamma_\ell$ is the eigenvalue of $\displaystyle \hat{D}_b ^{(\ell)}$ in the appropriate Fock space,
\begin{equation}
\hat{D}_b ^{(\ell)} \, |n\rangle \,=\, \gamma_\ell \, |n\rangle  \; . \label{DGAMMAL}
\end{equation}
For the consistency of the construction one can utilize Eqs. \eqref{NK-FINAL}, \eqref{QK}, to substitute the integral part operators in the definition \eqref{Db} obtaining
\begin{equation}
\hat{D}_{b} ^{(\ell)} \,=\, \frac{b-1}{2} \mathbb{I} + \sum^{b^\ell-1}_{j=1} C_j ^{(b^\ell)} \zeta_j^{\hat{n}} - b \sum^{b^{\ell+1}-1}_{j=1} C_j ^{(b^{\ell+1})} \zeta_j^{\hat{n}} \; . \label{DBEXPL1}
\end{equation}

\subsection{The $\gamma_{\ell}$'s via translation operators}
The classical method to calculate the coefficients $\gamma_{\ell}$ of the expansion of $n$ in the base $b$ can be expressed in terms of the recursive formula
\begin{equation}
\gamma_{\ell} \,=\, M_{\ell} - b \, M_{{\ell}+1} \;,\; \ell = 0,\ldots,q-1 \; . \label{GAMMAL}
\end{equation}
In Eq. \eqref{GAMMAL} $q$ $=$ $\displaystyle \left\lfloor \log_b n \right\rfloor + 1$ is the number of slots of the finite register required to describe $n$ in the base $b$, while the quotients $\displaystyle M_j \,=\, \left\lfloor \frac{n}{b^j} \right\rfloor$, $j = 0,\ldots,q$, can be calculated utilizing the explicit expression of the integral part of a rational number which is obtained from the result \eqref{NK-FINAL} replacing $\hat{n}$ with $n$ throughout. Notice the special values $M_0$ $=$ $n$ and $M_q$ $=$ 0, the latter corresponding to the end of the recursive calculations, i.e., to the result $\gamma_{q-1}$ $=$ $M_{q-1}$. For convenience, all computations are usually carried out in the base 10.  

In quantum terms, one defines the states $|\gamma_\ell\rangle$ $=$ $|M_\ell-W_\ell\rangle$ where $\displaystyle W_\ell$ $=$ $\displaystyle b \, M_{\ell+1}$ is the shift. With $\displaystyle W_\ell$ $\leq$ $\displaystyle M_\ell$ $\leq$ $n$, one applies Eq. (\ref{TCONT2}) obtaining the required translation
\begin{equation}
|\gamma_\ell\rangle \,=\, T_0 ^{W_\ell} |M_\ell\rangle \; , \label{GAMMALrec}
\end{equation}
which gives $\gamma_\ell$ the physical meaning of occupation numbers of shifted number states. For the correct application of the translation transformation, the state $|M_\ell\rangle$ in Eq. \eqref{GAMMALrec} is written in tensor product form using the expansion of $M_\ell$ in base 10.

No conceptual modifications are required in the definition of the translation of number states for the infinite register.

\section{Conclusions}
We have constructed two sets of quantum operators, the Hermitian multiboson operators $\displaystyle \hat{D}_{b} ^{(\ell)}$ and the shift operators $\displaystyle T_m, T_m^{\dagger}$; the latter are unitary on a region whose largest extension is obtained for $m$ $=$ 0. Both sets are characterized by the following property: their action on the number state $|n\rangle$ leads to the coefficients of the expansion of the natural number $n$ in any base $b$.

We have also shown that the coefficients of a natural number $n$ in any base have a twofold physical meaning: they are either the eigenvalues of the operators $\displaystyle \hat{D}_{b} ^{(\ell)}$ in Fock space or the occupation numbers of the shifted number states in the quantum version of the classical recursive solution of the base change problem.

We conjecture that both approaches are closely related to the construction of a quantum algorithm performing the number base change.

\end{document}